\documentstyle[aps]{revtex}

\begin{document}
\title{Are There Low-Lying Intruder States in $^8Be$?}

\author{M.S. Fayache$^{1,2}$, L. Zamick$^2$ and Y.Y. Sharon$^2$\\
(1) D\'{e}partement de Physique, Facult\'{e} des Sciences de Tunis\\
Tunis 1060, Tunisia\\
\noindent (2) Department of Physics and Astronomy, Rutgers University\\
        Piscataway, New Jersey 08855\\}
\date{\today}
\maketitle

\section{Introduction and Motivation}

In an $R$ matrix analysis of the $\beta^{\mp}$-delayed alpha spectra
from the decay of $^8Li$ and $^8B$ as measured by Alburger, Donavan
and Wilkinson \cite{adw}, Warburton \cite{war} made the
following statement in the abstract: ``It is found that satisfactory
fits are obtained without introducing intruder states below
26-$MeV$ excitations''. However, Barker has questioned this
\cite{bar1,bar2} by looking at the systematics of intruder states in
neighboring nuclei. He noted that the excitation energies of $0_2^+$
states in $^{16}O$, $^{12}C$ and $^{10}Be$ were respectively 6.05
$MeV$, 7.65 $MeV$ and 6.18 $MeV$. Why should there not then be an
intruder state in $^8Be$ around that energy?

In recent works \cite{fay1,fay2} the current authors and S. S. Sharma
allowed up to $2 \hbar \omega$ excitations in $^8Be$ and in $^{10}Be$,
and indeed $2p-2h$ intruder states were studied with some care in
$^{10}Be$. Using a simple quadrupole-quadrupole interaction $-\chi
Q\cdot Q$ with $\chi$=0.3615 $MeV/fm^4$ for $^{10}Be$ and $\hbar
\omega=45/A^{1/3}-25/A^{2/3}$. We found a $J=0^+$ intruder state at
9.7 $MeV$ excitation energy. This is higher than the experimental
value of 6.18 $MeV$, but it is in the ballpark. However, there are
other $J=0^+$ excited states below the intruder state found in the 
calculation. 

In a $0p$-shell calculation with the interaction $-\chi Q\cdot Q$, 
using a combination of the Wigner Supermultiplet theory \cite{wig}
characterized by the quantum numbers [$f_1 f_2 f_3$] and Elliott's
$SU(3)$ formula \cite{Elliott}, one can obtain the following
expression giving the energies of the various states: 

\[E(\lambda~\mu)=
\bar{\chi}\left[-4(\lambda^2+\mu^2+\lambda\mu+ 3
(\lambda+\mu))+3L(L+1)\right]\]

\noindent where 

\[\lambda=f_1-f_2,~~~\mu=f_2-f_3\]

\noindent and

\[\bar{\chi}=\chi \frac{5b^4}{32\pi}~~~~(b^2=\frac{\hbar}{m\omega})\]

The two $J=0^+$ states lying below the calculated intruder state in
$^{10}Be$, at least in the calculation, correspond to two degenerate
configurations [411] and [330]. Both of these have configurations
$L=1~S=1$ from which one can form the triplet configurations
$J=0^+,~1^+,~2^+$. Hence, besides the intruder state, we have the
above two $J=0^+$ states as candidates for the experimental $0_2^+$
state at 6.18 $MeV$.

As noted in the previous work \cite{fay1} if, in the $0p$-shell model
space we fit $\chi$ to get the energy of the lowest $2^+$ state in
$^{10}Be$ to be at the experimental value of 3.368 $MeV$
(18$\bar{\chi}$), then the two sets of triplets 
are at an excitation energy of 30 $\bar{\chi}$ which equals 5.61 $MeV$
-not far from the experimental value. There is however a problem -in a
$0p$-space calculation with $Q \cdot Q$, the lowest $2^+$ state is
two-fold degenerate, corresponding to $J=2^+~K=0$ and $J=2^+~K=2$. 

So it is by no means clear if the $0^+$ state in $^{10}Be$ at 6.18
$MeV$ is an intruder state. We will discuss this more in a later
section. It should be noted that in the previously mentioned
calculation \cite{fay2}, the energy of the intruder state is very
sensitive to the value of $\chi$, the strength of the $Q \cdot Q$
interaction. The energy of this intruder state drops down rapidly and
nearly linearly with increasing $\chi$. 

\section{Results}

In tables I, II and III we give results for the energies of $J=0^+$
and $2^+$ states in $^8Be$, in which up to $4 \hbar\omega$ excitations
are allowed relative to the basic configurations $(0s)^4(0p)^4$. The
different tables correspond to different interactions as follows:\\
(a) Quadrupole-Quadrupole: $V=-\chi Q \cdot Q$ with
$\chi=0.3467~MeV/fm^4$. \\
(b) $V=-\chi Q \cdot Q~+~xV_{s.o.}$ ($\chi$ as above and $x=1$).\\
(c) $V=V_{c}+xV_{s.o.}+yV_{t}$ ($x=1,~y=1$).\\

In the above, $s.o.$ stands for spin-orbit, $t$ for tensor and $c$ for
central. $V$ is a {\em two-body } interaction which for $x=1,~y=1$
gives a good fit to the Bonn A non-relativistic $G$ matrix
elements. This has been discussed extensively in previous references
\cite{fay1,ann}. 

In tables IV, V and VI we present results for isospin one $J=0^+$ and
$2^+$ states in $^{10}Be$ in which up to $2 \hbar \omega$ excitations
have been included. We have the same three interactions as above but
with $\chi=0.3615~MeV/fm^4$ in (a) and (b).

In all the tables we give the excitation energies of the $J=0^+$ and
$2^+$ states and the percent probability that there are no excitations
beyond the basic configuration ($0 \hbar \omega$) and the percentage
of $2 \hbar \omega$ excitations (as well as $4 \hbar \omega$
excitations for $^8Be$).

Note that for interaction (a) the respective percentages for the
ground state of $^8Be$ (see table I) are 62.8\%, 25.7\% and
11.5\%: there is considerable mixing. Thus we should not forget, when
we discuss the question ``where are the intruder states?'', that there
is considerable admixing of $2 \hbar \omega$ and $4 \hbar \omega$
excitations {\em in the ground state}. Note that the ground state
configuration does not change very much for the three interactions
that are considered here. For example, as seen in table III, the
corresponding percentages for the ($x,y$) interaction are 62.2\%,
26.2\% and 11.6\%. 

By looking at these tables, it is not too difficult to see at what
energies the intruder states set in. One sees a sharp drop in
the $0 \hbar \omega$ occupancy. For example in table I, whereas the 
$0 \hbar \omega$ percentage for the 18.7 $MeV$ and 20.2 $MeV$ states
are respectively 93.9\% and 94.6\%, for the next state at 26.5 $MeV$
the percentage drops to 29.4\% -also the next four states listed have
very low $0 \hbar \omega$ percentages and are therefore intruders. 

Let us somewhat arbitrarily define an intruder state as one for which
the $0 \hbar \omega$ percentage is less than 50\%. By this criterion,
and for the three interactions discussed here, the lowest $J=0^+$
intruder states in $^8Be$ are at 26.5 $MeV$, 26.5 $MeV$ and 28.7 $MeV$
(see tables I,II and III). The $J=2^+$ intruder states are at 27.5
$MeV$, 27.5 $MeV$ and 33.7 $MeV$. Note that up to 4$\hbar \omega$
excitations were allowed in these calculations. These energies are
very high and would argue against the suggestion by Barker that there
are low-lying intruder states in $^8Be$. 

What about $^{10}Be$? Remember that in this nucleus we only include 
up to $2 \hbar \omega$ excitations. For the three interactions
considered, the lowest $J=0^+~T=1$ intruder states are at 9.7 $MeV$, 11.4
$MeV$ and 31.0 $MeV$. The `anomalous' behavior for the last value
(31.5 $MeV$ for the ($x,y$) interaction) will be discussed in a later
section.  

Note that when a spin-orbit is added to $Q \cdot
Q$, the energy of the intruder state goes up $e.g.$ 11.4 $MeV$ $vs$ 9.7
$MeV$. The lowest-lying $J=2^+~T=1$ intruder states are at 11.9 $MeV$,
13.8 $MeV$ and 33.4 $MeV$. The energy of the non-intruder ($L=1~S=1$)
$J=0^+,~1^+,~2^+$ triplet also goes up as can be seen from tables IV and V.

For the two $Q \cdot Q$ interactions, the energies of the intruder
states in $^{10}Be$ are much lower than in $^8Be$. This conclusion
still holds if we were to use $^8Be$ energies calculated in
(0+2)$\hbar \omega$ configuration space -see table VII. This would
indicate that even if we do find low-lying intruder states in
$^{10}Be$, such a finding in itself is not proof that they are also
present in $^8Be$. Indeed, our calculations would dispute this
claim. 

\section{ $(0+2) \hbar \omega$ $vs$ $(0+2+4) \hbar \omega$
Calculations for $^8Be$}

In table VII we show the results for the energy of the first intruder
state in $^8Be$ in calculations in which only up to $2 \hbar \omega$
excitations are included, and compare them with the corresponding
results for up to $4 \hbar \omega$. For interactions (a) and (b), the
value of $\chi$ was changed to 0.4033 $MeV/fm^4$ in order that the
energy of the $2_1^+$ state come close to experiment. In more detail,
we have to rescale $\chi$ depending on the model space in order to
get the $2_1^+$ state at the right energy. In general, the more
$np-nh$ configurations we include the smaller $\chi$ is.

We see that in the larger-space calculation $(0+2+4) \hbar
\omega$, the energies of the lowest intruder states in most cases come
down about 5 $MeV$ relative to the $(0+2) \hbar \omega$ calculation. The
excitation energies are still quite high, however, all being above 25
$MeV$. One possible reason for the difference between the results of
the two calculations is that in the $(0+2)\hbar \omega$ calculation
there is level repulsion between the $0 \hbar \omega$ and the $2 \hbar
\omega$ configurations, and that the $4 \hbar \omega$ configurations
are needed to repel the $2 \hbar \omega$ states back down. 

\section{The first excited $J=0^+$ state of $^{10}Be$}

Is the first excited $J=0^+$ state in $^{10}Be$ an intruder state or
is it dominantly of the $(0s)^4(0p)^6$ configuration? Experimentally,
very few states have been identified in $^{10}Be$. The known
positive-parity states are as follows \cite{ajz}:

\begin{tabbing}
\= \hspace{1.in} \= $J^{\pi}$\hspace{0.3in} \= $E_x(MeV)$  \\
\>		 \> $0_1^+$                 \> 0.000 \\
\>               \> $2_1^+$                 \> 3.368 \\
\>               \> $2_2^+$                 \> 5.959 \\
\>               \> $0^+$                   \> 6.179 \\
\>               \> $2^+$                   \> 7.542 \\
\>               \> ($2^+$)                 \> 9.400 \\
\end{tabbing}

In the $(0s)^4(0p)^6$ calculation with a $Q \cdot Q$ interaction, the
lowest $2^+$ state at 18$\bar{\chi}$ is doubly degenerate and
corresponds to $K=0$ and $K=2$ members of the [42]
configuration. There are two degenerate ($L=1~S=1$) configurations at
30$\bar{\chi}$ with supermultiplet configurations [330] and
[411]. From $L=1~S=1$ one can form a triplet of states with
$J=0^+,~1^+,~2^+$. If we choose $\bar{\chi}$ by getting the $2_1^+$
state correct at 3.368 $MeV$, then the two $L=1~S=1$ triplets would
be at $30/18\times 3.36~MeV=5.61~MeV$. However, there should be a {\em
triplet} of states. In more detailed calculations, as the spin-orbit
interaction is added to the $Q \cdot Q$ interaction, the triplet
degeneracy gets removed with the ordering
$E_{2^+}<E_{1^+}<E_{0^+}$. As seen in tabel IV, the $J=0^+$ and $2^+$
states of $^{10}Be$ at 3.7 $MeV$ and 7.3 $MeV$ are degenerate with a
pure $Q \cdot Q$ interaction. This is also true for $J=1^+$. In table
V, however, when the spin-orbit interaction is added to $Q \cdot Q$,
we find that whereas the $0^+_2$ is at 8.0 $MeV$, the $2^+_3$ state is
at 6.8 $MeV$.

Hence if the $0^+$ state at 6.179 $MeV$ were dominantly an $L=1~S=1$
non-intruder state, one would expect a $J=1^+$ and a $J=2^+$ state at
lower energies. Thus far no $J=1^+$ level has been seen in $^{10}Be$
but this is undoubtedly due to the lack of experimental research on
this target. Now there {\em is} a lower $2^+$ state at 5.959
$MeV$. This could be a member of the $L=1~S=1$ triplet {\em or} it
could be the $K=2$ state of the [42] configuration. 

Hence, one possible scenario is that indeed the $2^+_2$ state is
dominantly of the [42] configuration and the $J=0_2^+$ state is a
singlet. This would support the idea that the $J=0^+_2$ state is an
intruder state. The second scenario has the $J=2_2^+$ state being
dominantly an $L=1~S=1$ state for which the $J=1^+$ member has somehow
not been found. This would be in support of the idea that the $0_2^+$
state is {\em not} an intruder state. 

Let us look in detail at tables IV, V and VI which show where the
energies of the intruder states are in a (0+2)$\hbar \omega$
calculation. For the $Q \cdot Q$ interaction (with
$\chi=0.3615~MeV/fm^4$), the lowest $J=0^+$ intruder state is at 9.7
$MeV$ and the lowest $J=2^+$ intruder state is at 11.9 $MeV$. These
energies are {\em much lower} than the corresponding intruder state
energies for $^8Be$. This in itself is enough to tell us that the
presence of a low-energy intruder state in $^{10}Be$ does not imply
that there should be a low energy intruder state in $^8Be$. Note that
the intruder states in this model space and with this interaction have
100\% `2$\hbar \omega$' configurations. This has been noted and
discussed in \cite{fay2} and is due to the fact that the $Q \cdot Q$
interaction cannot excite two nucleons from the $N$ shell to the $N\pm
1$ shell. 

Still, in table IV, there are two $J=0^+$ states (below the intruder
state) at 3.7 $MeV$ and 7.3 $MeV$. Even in this large-space
calculation, these are members of degenerate $L=1~S=1$ triplets
$J=0^+,~1^+,~2^+$. Indeed, if we look down the table, we see the 3.7
$MeV$ and 7.3 $MeV$ values in the $J=2^+$ column.

In table V, when we add the spin-orbit interaction to $Q \cdot Q$, the
energies of the $0^+_2$ and $0^+_3$ states go up, but so does the
energy of the $J=0^+_4$ intruder state. The energies of the $0^+_2$,
$0^+_3$ and $0^+_4$ (intruder) states in table IV are 3.7, 7.3 and 9.7
$MeV$; in table V, with the added spin-orbit interaction they are 8.0,
9.6 and 11.4 $MeV$.

In table VI we show results of an up-to-2$\hbar \omega$ calculation
with the realistic interaction. Here, we see a drastically different
behavior for the intruder state energy in $^{10}Be$. The lowest
$J=0^+$ intruder state is at 31.0 $MeV$, and the lowest $J=2^+$
intruder state is at 33.4 $MeV$ (recall our operational definition -an
intruder state has less than 50\% of the 0$\hbar \omega$
configuration). For the $Q \cdot Q$ interaction, in contrast, the
intruder state was at a much lower energy. A possible explanation is
that for the ($x,y$) interaction, unlike $Q \cdot Q$, one {\em does
have} large off-diagonal matrix elements in which two nucleons are
excited from $N$ to $N\pm 1$ $e.g.$ from $0p$ to $1s-0d$. This will
cause a large level repulsion between the 0$\hbar \omega$ and the 
2$\hbar \omega$ configurations and drive them far apart. Presumably,
if we included 4$\hbar \omega$ configurations into the model space,
they would push the 2$\hbar \omega$ configurations back down to near
their unperturbed positions. 

Thus, the problem is rather difficult to sort out theoretically, so we
can at best suggest that more experiments be done on $^{10}Be$. For
example, the $B(E2)$ to the $2_2^+$ state would be useful. There
should be a much larger B(E2) to the $L=2~K=2$ member of a [42]
configuration than to an ($L=1~S=1$) state. We also predict a
substantial $B(M1)\uparrow$ to the first $J=1^+~T=1$ state in
$^{10}Be$. Whereas with a pure $Q \cdot Q$ interaction the $B(M1)$ to 
this state would be zero, the presence of a spin-orbit interaction
will `light up' the $1^+_1$ state in $^{10}Be$. The $J=1^+$ should be
seen. 

\section{Conclusions}

Because of the important implications to astrophysics of the $^8Be$
nucleus, we feel that Barker's suggestion of looking for intruder 
states in this and neighboring nuclei is well founded. In all our
calculations, the positive-parity intruder states in $^8Be$ come at a
very high excitation energy -greater than 26 $MeV$, thus supporting
the statement by E. Warburton in the abstract of his 1986 work
\cite{war}. However, because of significant differences between the
realistic and the $Q \cdot Q$ interactions for the predicted energies
of intruder states in $^{10}Be$, and because of the possibility of
low-lying non-intruder states $e.g.$ $L=1~S=1$ triplets, we cannot
determine with certainty whether or not the first excited state in 
$^{10}Be$ is an intruder state. However, even if it is, this does not
mean that there has to be a low-energy intruder state in
$^8Be$. Indeed, our $Q \cdot Q$ calculations clearly contradict this
claim, and in $^8Be$, our realistic-interaction calculations lead to
the same conclusion.

\section{Acknowledgements}

We are extremely grateful to Ian Towner for suggesting this
problem. This work was supported by a Department of Energy Grant
No. DE-FG02-95ER40940 and by a Stockton College Distinguished Faculty
research grant. M.S. Fayache kindly acknowledges travel support from
the Laboratoire de la Physique de la Mati\`{e}re Condens\'{e}e at the
Universit\'{e} de Tunis, Tunisia.

\nopagebreak

\pagebreak

\begin{table}
\caption{$J=0^+$ and $2^+$ states in $^8Be$ for the interaction $-\chi
Q\cdot Q$ with $\chi=0.3457~MeV/fm^4$ with up to $4\hbar\omega$
excitations allowed. The percentage of $0\hbar\omega$, $2\hbar\omega$
and $4\hbar\omega$ occupancies are given, as well as the $B(E2)(0^+_1
\rightarrow 2^+_i$.}
\begin{tabular}{ccccc}
\multicolumn{5}{c}{(a) $J=0^+~T=0$ States}\\
{$E_{exc}(MeV)$} & $0~\hbar\omega$ & $2~\hbar\omega$ & $4~\hbar\omega$
& \\
\tableline
 0.0 & 62.8 & 25.7 & 11.5 & \\
11.9 & 82.2 & 11.6 &  6.2 & \\
16.6 & 94.0 &  2.2 &  3.8 & \\
18.7 & 93.9 &  2.7 &  3.4 & \\
20.3 & 94.6 &  2.3 &  3.2 & \\
26.5 & 29.4 & 49.7 & 20.9 & \\
29.9 &  4.1 & 75.8 & 20.1 & \\
32.4 &  0.0 & 85.4 & 14.6 & \\
34.5 &  0.0 & 86.2 & 13.9 & \\
36.4 & 14.8 & 69.3 & 15.8 & \\
\tableline
\multicolumn{5}{c}{(b) $J=2^+~T=0$ States}\\
{$E_{exc}(MeV)$} & $0~\hbar\omega$ & $2~\hbar\omega$ & $4~\hbar\omega$
& $B(E2)_{0^+_1 \rightarrow 2^+_i}~(e^2fm^4)$\\
\tableline
 3.1 & 64.3 & 25.0 & 10.7 & 67.2 \\
11.9 & 82.2 & 11.6 &  6.2 &  0.0 \\
14.2 & 85.1 &  9.6 &  5.3 &  0.0 \\
16.7 & 87.6 &  8.0 &  4.4 &  0.0 \\
16.7 & 92.9 &  3.2 &  3.9 &  0.0 \\
18.7 & 93.9 &  2.7 &  3.4 &  0.0 \\
18.7 & 93.9 &  2.7 &  3.4 &  0.0 \\
20.3 & 94.6 &  2.3 &  3.2 &  0.0 \\
27.5 & 29.9 & 49.6 & 20.5 & 15.2 \\
30.4 &  0.0 & 78.4 & 21.6 &  0.0 \\
32.0 &  1.2 & 79.1 & 19.7 &  1.7 \\
32.4 &  0.0 & 85.4 & 14.6 &  0.0 \\
34.2 &  0.1 & 82.5 & 17.4 &  0.0 \\
34.5 &  0.0 & 86.1 & 13.9 &  0.0 \\
36.2 & 11.6 & 73.7 & 14.7 &  0.0 \\
\end{tabular}
\end{table}

\begin{table}
\caption{Same as Table I but for the interaction $-\chi Q\cdot Q + x
V_{s.o.}$ with $\chi=0.3457~MeV/fm^4$ and $x=1$.}
\begin{tabular}{ccccc}
\multicolumn{5}{c}{(a) $J=0^+~T=0$ States}\\
{$E_{exc}(MeV)$} & $0~\hbar\omega$ & $2~\hbar\omega$ & $4~\hbar\omega$
& \\
\tableline
 0.0 & 65.1 & 24.0 & 10.9 & \\
12.8 & 83.6 & 10.3 &  6.1 & \\
16.4 & 89.7 &  6.0 &  4.3 & \\
21.9 & 91.7 &  4.6 &  3.7 & \\
26.4 & 69.3 & 21.3 &  9.4 & \\
26.5 & 40.7 & 44.0 & 15.3 & \\
29.9 &  3.4 & 77.4 & 19.2 & \\
32.1 &  0.0 & 86.6 & 13.4 & \\
37.3 &  0.0 & 85.6 & 14.3 & \\
38.4 & 18.2 & 66.2 & 15.6 & \\
\tableline
\multicolumn{5}{c}{(b) $J=2^+~T=0$ States}\\
{$E_{exc}(MeV)$} & $0~\hbar\omega$ & $2~\hbar\omega$ & $4~\hbar\omega$
& $B(E2)_{0^+_1 \rightarrow 2^+_i}~(e^2fm^4)$\\
\tableline
 3.1 & 66.7 & 23.3 & 10.1 & 63.4 \\
10.2 & 85.8 &  8.8 &  5.4 &  0.4 \\
13.2 & 88.2 &  7.2 &  4.6 &  0.9 \\
16.2 & 91.9 &  4.2 &  3.9 &  0.0 \\
17.7 & 86.4 &  8.9 &  4.7 &  0.2 \\
19.6 & 88.3 &  7.4 &  4.3 &  0.0 \\
21.6 & 84.8 & 10.3 &  4.9 &  0.1 \\
22.2 & 91.0 &  5.1 &  3.8 &  0.0 \\
27.5 & 27.8 & 53.1 & 19.1 & 14.5 \\
30.9 &  0.9 & 78.0 & 21.0 &  0.0 \\
31.9 &  1.1 & 80.2 & 18.7 &  1.6 \\
32.4 &  0.0 & 86.2 & 13.8 &  0.0 \\
34.3 &  0.2 & 85.7 & 14.0 &  0.0 \\
34.6 &  1.2 & 83.8 & 15.1 &  0.1 \\
35.2 & 11.4 & 74.0 & 14.6 &  0.1 \\
\end{tabular}
\end{table}

\begin{table}
\caption{Same as Table I but for the realistic ($x,y$) interaction 
with $x=1$ and $y=1$.}
\begin{tabular}{ccccc}
\multicolumn{5}{c}{(a) $J=0^+~T=0$ States}\\
{$E_{exc}(MeV)$} & $0~\hbar\omega$ & $2~\hbar\omega$ & $4~\hbar\omega$
& \\
\tableline
 0.0 & 62.2 & 26.2 & 11.6 & \\
22.8 & 66.5 & 23.6 &  9.9 & \\
28.7 &  6.5 & 71.0 & 22.5 & \\
30.3 & 66.5 & 23.0 & 10.5 & \\
35.3 & 67.5 & 22.4 & 10.1 & \\
39.4 &  7.3 & 73.4 & 19.3 & \\
43.5 & 56.3 & 31.4 & 12.3 & \\
47.6 &  8.8 & 70.5 & 20.7 & \\
49.5 &  2.3 & 76.7 & 21.6 & \\
50.1 &  3.3 & 75.7 & 21.0 & \\
\tableline
\multicolumn{5}{c}{(b) $J=2^+~T=0$ States}\\
{$E_{exc}(MeV)$} & $0~\hbar\omega$ & $2~\hbar\omega$ & $4~\hbar\omega$
& $B(E2)_{0^+_1 \rightarrow 2^+_i}~(e^2fm^4)$\\
\tableline
 5.4 & 62.2 & 26.6 & 11.1 & 31.1 \\
19.5 & 70.0 & 20.4 &  9.6 &  0.0 \\
21.5 & 69.5 & 20.2 & 10.3 &  0.1 \\
26.2 & 69.7 & 20.5 &  9.8 &  0.4 \\
30.4 & 70.2 & 20.9 &  8.9 &  0.0 \\
31.0 & 56.7 & 30.9 & 12.6 &  1.7 \\
33.7 & 13.5 & 65.7 & 20.8 &  3.7 \\
35.1 & 71.3 & 19.7 &  9.0 &  0.0 \\
38.2 & 67.7 & 22.4 &  9.8 &  0.0 \\
41.6 &  9.0 & 68.8 & 22.2 &  1.3 \\
45.0 &  1.0 & 79.7 & 19.3 &  0.1 \\
45.9 &  2.9 & 77.9 & 19.2 &  2.4 \\
46.3 &  3.2 & 76.7 & 20.1 &  1.3 \\
47.3 &  0.3 & 79.5 & 20.2 &  0.0 \\
48.4 &  1.5 & 79.8 & 18.6 &  0.0 \\
\end{tabular}
\end{table}

\begin{table}
\caption{$J=0^+$ and $2^+$ states in $^{10}Be$ for the interaction $-\chi
Q\cdot Q$ with $\chi=0.3615~MeV/fm^4$ with up to $2\hbar\omega$
excitations allowed. The percentage of $0\hbar\omega$ 
and $2\hbar\omega$ occupancies are given, as well as the $B(E2)(0^+_1
\rightarrow 2^+_i$.}
\begin{tabular}{cccc}
 & \multicolumn{2}{c}{(a) $J=0^+~T=1$ States} & \\
{$E_{exc}(MeV)$} & $0~\hbar\omega$ & $2~\hbar\omega$ & \\
\tableline
 0.0 & 81.8 & 18.2 & \\
 3.7 & 81.0 & 19.0 & \\
 7.3 & 93.6 &  6.4 & \\
 9.7 &  0.0 &100.0 & \\
12.1 & 92.9 &  7.1 & \\
12.1 & 92.9 &  7.1 & \\
13.9 & 93.1 &  6.9 & \\
17.7 & 98.9 &  1.1 & \\
22.1 &  0.0 &100.0 & \\
22.9 &  0.0 &100.0 & \\
\tableline
 & \multicolumn{2}{c}{(b) $J=2^+~T=1$ States} & \\
{$E_{exc}(MeV)$} & $0~\hbar\omega$ & $2~\hbar\omega$ 
& $B(E2)_{0^+_1 \rightarrow 2^+_i}~(e^2fm^4)$\\
\tableline
 2.2 & 81.3 & 18.7 &  5.0 \\
 3.4 & 83.4 & 16.6 & 47.2 \\
 3.7 & 81.0 & 19.0 &  0.0 \\
 7.3 & 93.6 &  6.4 &  0.0 \\
 9.2 & 82.9 & 17.1 &  0.0 \\
10.9 & 91.9 &  8.1 &  0.0 \\
11.9 &  0.0 &100.0 &  0.0 \\
12.1 & 92.9 &  7.1 &  0.0 \\
12.1 & 92.9 &  7.1 &  0.0 \\
12.1 & 92.9 &  7.1 &  0.0 \\
13.9 & 93.1 &  6.9 &  0.2 \\
13.9 & 93.1 &  6.9 &  0.0 \\
13.9 & 93.1 &  6.9 &  0.0 \\
17.7 & 98.9 &  1.1 &  0.0 \\
22.1 &  0.0 &100.0 &  0.0 \\
\end{tabular}
\end{table}

\begin{table}
\caption{Same as Table IV but for the interaction $-\chi Q\cdot Q + x
V_{s.o.}$ with $\chi=0.3615~MeV/fm^4$ and $x=1$.}
\begin{tabular}{cccc}
 & \multicolumn{2}{c}{(a) $J=0^+~T=1$ States} & \\
{$E_{exc}(MeV)$} & $0~\hbar\omega$ & $2~\hbar\omega$ & \\
\tableline
 0.0 & 85.6 & 14.4 & \\
 8.0 & 80.8 & 19.2 & \\
 9.6 & 92.0 &  8.0 & \\
11.4 &  0.0 &100.0 & \\
12.1 & 91.5 &  8.5 & \\
16.4 & 90.6 &  9.4 & \\
19.7 & 90.5 &  9.5 & \\
23.1 & 88.7 & 11.3 & \\
24.0 &  0.0 &100.0 & \\
26.1 &  0.0 &100.0 &\\
\tableline
 & \multicolumn{2}{c}{(b) $J=2^+~T=1$ States} & \\
{$E_{exc}(MeV)$} & $0~\hbar\omega$ & $2~\hbar\omega$ 
& $B(E2)_{0^+_1 \rightarrow 2^+_i}~(e^2fm^4)$\\
\tableline
 3.0 & 85.5 & 14.5 & 40.1 \\
 4.6 & 83.7 & 16.3 &  3.4 \\
 6.8 & 90.8 &  9.2 &  0.3 \\
 7.8 & 83.5 & 16.5 &  3.7 \\
11.8 & 84.8 & 15.2 &  0.1 \\
13.0 & 91.2 &  8.8 &  0.1 \\
13.8 &  0.0 &100.0 &  0.0 \\
14.1 & 90.9 &  9.1 &  0.0 \\
14.8 & 90.9 &  9.1 &  0.0 \\
15.5 & 90.3 &  9.7 &  0.0 \\
17.2 & 90.0 & 10.0 &  0.1 \\
17.2 & 88.0 & 12.0 &  0.0 \\
18.2 & 90.3 &  9.7 &  0.1 \\
21.2 & 89.0 & 11.0 &  0.0 \\
23.0 & 52.8 & 47.3 &  0.0 \\
\end{tabular}
\end{table}

\begin{table}
\caption{Same as Table IV but for the realistic ($x,y$) interaction 
with $x=1$ and $y=1$.}
\begin{tabular}{cccc}
 & \multicolumn{2}{c}{(a) $J=0^+~T=1$ States} & \\
{$E_{exc}(MeV)$} & $0~\hbar\omega$ & $2~\hbar\omega$ & \\
\tableline
 0.0 & 73.3 & 26.7 & \\
 8.7 & 74.4 & 25.6 & \\
12.0 & 74.7 & 25.3 & \\
21.1 & 76.5 & 23.5 & \\
23.7 & 77.5 & 22.5 & \\
31.0 & 49.3 & 50.7 & \\
31.5 & 25.4 & 74.6 & \\
34.5 &  5.8 & 94.2 & \\
37.6 &  0.6 & 99.4 & \\
39.7 & 74.1 & 25.9 & \\
\tableline
 & \multicolumn{2}{c}{(b) $J=2^+~T=1$ States} & \\
{$E_{exc}(MeV)$} & $0~\hbar\omega$ & $2~\hbar\omega$ 
& $B(E2)_{0^+_1 \rightarrow 2^+_i}~(e^2fm^4)$\\
\tableline
 4.6 & 73.5 & 26.5 & 19.7 \\
 5.2 & 73.9 & 26.1 &  3.2 \\
 9.2 & 73.7 & 26.3 &  1.5 \\
10.1 & 75.8 & 24.2 &  0.0 \\
17.4 & 74.5 & 25.5 &  0.0 \\
19.7 & 75.7 & 24.3 &  0.1 \\
20.2 & 77.0 & 23.0 &  0.0 \\
22.1 & 76.9 & 23.1 &  0.2 \\
22.9 & 77.1 & 22.9 &  0.0 \\
23.7 & 77.2 & 22.8 &  0.0 \\
27.2 & 76.8 & 23.2 &  0.0 \\
29.0 & 76.9 & 23.1 &  0.2 \\
32.5 & 76.9 & 23.1 &  0.0 \\
33.4 &  0.3 & 99.7 &  0.0 \\
35.5 & 71.7 & 28.3 &  0.2 \\
\end{tabular}
\end{table}

\begin{table}
\caption{Excitation energies (in $MeV$) of the first $J=0^+$ and $2^+$
intruder states in $^8Be$: a comparison of up to $2\hbar\omega$ and 
up to $4\hbar\omega$ calculations for the three interactions.}
\begin{tabular}{cccc}
 & $Q \cdot Q$ & $Q \cdot Q~+~xV_{s.o.}$ & ($x,y$)=(1,1) \\
\tableline
 & \multicolumn{3}{c}{$J=0^+~T=0$ States}\\
$2\hbar\omega$ & 32.1 & 30.1 & 33.8 \\
$4\hbar\omega$ & 26.5 & 26.5 & 28.7 \\
\tableline
 & \multicolumn{3}{c}{$J=2^+~T=0$ States}\\
$2\hbar\omega$ & 31.5 & 30.9 & 36.6 \\
$4\hbar\omega$ & 27.5 & 27.5 & 33.7 \\
\end{tabular}
\end{table}

\end{document}